# Границы магнитных доменов как мезоскопические объекты

О. П. Мартыненко, В. В. Махро, И.Г. Махро

Братский государственный технический университет

В первой части статьи приводятся основные положения микромагнитной и квантовой теорий доменных структур в различных типах магнетиков. Формулируются модели, используемые в настоящей работе, и вводятся основные уравнения теории. Обсуждается характер взаимодействия границ магнитных доменов с магнитными неоднородностями и устанавливается взаимосвязь между макроскопическими характерстиками процессов перемагничивания и микроскопическими динамическими параметрами смещения границ магнитных доменов.

Здесь же обсуждаются основные механизмы влияния процессов термоактивационного и туннельного срывов на макроскопические характеристики процессов перемагничивания.

Во второй части рассматриваются вопросы, касающиеся механизма туннельного срыва в ферромагнетиках и слабых ферромагнетиках. Рассматриваются результаты, полученные к настоящему времени в экспериментах других авторов по изучению туннелирования границ, и обсуждаются теоретические методы изучения механизма туннельного срыва для доменных границ.

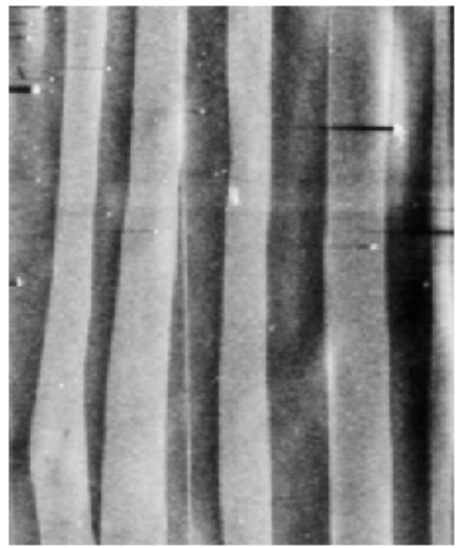

Рис. 1. Типичная доменная структура в мягком ферромагнетике (аморфный сплав $Fe_{64}Co_{21}B_{15}$, взято из [28]).

# 1 Уравнения движения намагниченности и динамики границ

Для удобства дальнейшего чтения в этом разделе мы вначале приводим краткую сводку основных положений микромагнитной теории доменных структур и доменных границ. Более полные сведения по этому вопросу можно найти, например, в монографиях [1] и [2].

## 1.1 Доменные границы в ферромагнетиках

Предположение о неоднородном распределении намагниченности в ферромагнетике было введено Вейссом [3]. В размагниченном состоянии ферромагнетик спонтанно разбивается на области, в пределах которых направление намагниченности остается неизменным (домены), тогда как может различаться для соседних областей. Возникающая при этом конфигурация намагниченности является минимизирующей для функционала полной энергии.

В пограничных областях намагниченность должна непрерывно изменяться от одного выделенного направления (в пределах одного домена), к другому — сформировавшемуся в пределах соседнего домена. Область, в пределах которой происходит это изменение направления, и называется доменной границей (стенкой).

Впервые понятие доменной стенки было введено Блохом [4] в 1932 году, он же выполнил первоначальные исследования влияния смещения доменных границ на формирование кривой намагничивания. Однако интенсивное изучение этих объектов продолжается вплоть до настоящего времени. Причины такого интереса понятны, если учесть, что доменные границы не только играют ключевую роль в динамике процессов перемагничивания, но и служат идеальными модельными объектами для многих современных разделов физики, таких как теория солитонов, или мезоскопическая физика.

Наиболее естественно представление о доменной структуре и доменных границах возникает в рамках так называемой микромагнитной теории, являющейся по сути континуальным феноменологическим описанием магнитных явлений. В теории микромагнетизма постулируется существование в ферромагнетике спонтанной намагниченности, описываемой векторным полем $\mathbf{M}(\mathbf{r},t)$ [5],[7],[8]. При этом часто предполагается, что модуль вектора намагниченности $M_0$ не зависит от пространственных координат, в силу чего можно писать $\mathbf{M} = M_0(\mathbf{r})\,\alpha(\mathbf{r})$, $\sum_{i=1}^{3}\alpha_i^2 = 1$. Равновесное распределение намагниченности определяется вариационным принципом: истинное распределение намагниченности дается минимизацией функционала полной свободной энергии кристалла. Последняя включает в себя: обменную энергию с плотностью

$$w_e = \frac{1}{2}f(M^2) + \frac{1}{2}\beta_{ij}(M^2)\frac{\partial \mathbf{M}}{\partial x_i}\frac{\partial \mathbf{M}}{\partial x_j},$$

причем структура тензора $\beta_{ij}$ в этом выражении определяется симметрией кристаллической решетки [1]; энергию анизотропии, которая описывает взаимодействие между анизотропной кристаллической решеткой и намагниченностью, и плотность которой дается для кубических кристаллов выражением

$$w_a = \beta_1(\alpha_1^2\alpha_2^2 + \alpha_1^2\alpha_3^2 + \alpha_2^2\alpha_3^2) + \beta_2\alpha_1^2\alpha_2^2\alpha_3^2 + ...,$$

а для гексагональных кристаллов

$$w_a = \beta_1\left(\alpha_1^2 + \alpha_2^2\right) + \beta_2\left(\alpha_1^2 + \alpha_2^2\right)^2 + ...$$

(преимущественные направления, определяемые этими выражениями, называются осями легкого намагничивания), энергию магнитного дипольного взаимодействия, плотность которой можно взять в форме[2]

$$w_m = -\frac{1}{2}\mathbf{H}_m\mathbf{M},$$

и, наконец, энергию взаимодействия намагниченности с внешним полем с плотностью

$$w_H = -\mathbf{H}_0^{ext}\mathbf{M}.$$

Полная макроскопическая энергия ферромагнетика во внешнем магнитном поле будет, таким образом,

$$W = \int dV(w_e + w_a + w_m + w_H) \qquad (1)$$

Варьирование (1)

$$\delta W = -\int dV\,\mathbf{H}_{eff}\delta\mathbf{M}, \qquad (2)$$

где $\mathbf{H}_{eff}$ - эффективное поле, определяемое как

$$\mathbf{H}_{eff} = -\frac{\partial W}{\partial \mathbf{M}} + \frac{\partial}{\partial x_i}\frac{\partial w}{\partial \frac{\partial \mathbf{M}}{\partial x_i}} + \mathbf{H}_0^{ext} + \mathbf{H}_m,$$

---

[1]Для кубических кристаллов квадратичная форма обменной энергии изотропна и плотность энергии имеет в этом случае вид

$$w_e = A\sum_{i,k=1}^{3}\left(\frac{\partial \alpha_i}{\partial x_k}\right)^2,$$

где $A$ - константа обменного взаимодействия.

[2]Поле $\mathbf{H}_m$ определяется решением системы уравнений магнитостатики $rot\mathbf{H}_m = 0, div(\mathbf{H}_m + 4\pi\mathbf{M}) = 0$.

приводит к уравнениям, предложенным Ландау и Лифшицем [9], [11]

$$\frac{\partial}{\partial t}\mathbf{M} = -\gamma \mathbf{M} \times \mathbf{H}_{eff} + \frac{\lambda}{M^2}\mathbf{M} \times (\mathbf{M} \times \mathbf{H}) \quad (3)$$

($\gamma = 2\mu_B/\hbar$, $\hbar$ - постоянная Планка, $\mu_B$ - магнетон Бора, $\lambda$ - константа диссипации).

В отсутствии диссипации (3) переходит в

$$\frac{\partial}{\partial t}\mathbf{M} = -\gamma \mathbf{M} \times \mathbf{H} \quad (4)$$

В линейном приближении (4) описывает спиновые волны в ферромагнетике. В главе 5 мы обсудим соответствующие решения линеаризованных уравнений применительно к поверхностным волнам на доменных границах. Однако сейчас мы сосредоточимся несколько подробнее на точных решениях уравнения Ландау — Лифшица. Возможны два типа нелинейных решений в зависимости от выбора граничных условий. Первому типу, так называемым динамическим солитонам, соответствует выбор

$$\mathbf{M}(\pm\infty) = \mathbf{M}_0, \quad (5)$$

граничные условия вида

$$\mathbf{M}(+\infty) = \mathbf{M}_0, \ \mathbf{M}(-\infty) = -\mathbf{M}_0 \quad (6)$$

определяют топологические солитоны — доменные границы[3]. Действительно, в сферической системе координат

$$M_z = M_0 \cos\theta, \ M_x = M_0 \sin\theta \cos\varphi, \ M_y = M_0 \sin\theta \sin\varphi \quad (7)$$

для кристалла ромбической сингонии с энергией анизотропии

$$\begin{aligned} w_a &= -\frac{1}{2}\beta M_z^2 + \frac{1}{2}\beta_1 M_x^2 = \quad (8) \\ &= \frac{1}{2}M_0^2 \sin^2\theta \left(1 + \epsilon \cos^2\varphi\right) - \frac{1}{2}\beta M_0^2 \end{aligned}$$

полная энергия записывается как

$$w = -\frac{1}{2}\beta M_0^2 + \frac{1}{2}\beta M_0^2 \sin^2\theta \left(1 + \epsilon \cos^2\varphi\right) + \frac{1}{2}\alpha M_0^2 \left[(\nabla\theta)^2 + \sin^2\theta (\nabla\varphi)^2\right], \quad (9)$$

---
[3]Часто в качестве синонима используется термин "кинки".

а уравнения (5) соответственно ([2], [12])

$$\frac{M_0}{\gamma}\frac{\partial}{\partial t}\cos\theta = \frac{\delta W}{\delta\varphi}, \ \frac{M_0}{\gamma}\frac{\partial\varphi}{\partial t} = -\frac{\delta W}{\delta\cos\theta}. \quad (10)$$

Точное решение (9), соответствующее граничному условию (6) (доменной границе), было найдено Уокером [13] в форме

$$\cos\theta = -\tanh\left(\frac{x - vt}{D(\varphi)}\right), \quad (11)$$

причем

$$\epsilon \sin\varphi \cos\varphi = \frac{v}{D\gamma\beta M_0}\left(1 + \epsilon \cos^2\varphi\right), \quad (12)$$

и $D = \sqrt{\alpha/\beta}$, $v$ - скорость доменной границы. Для покоящейся границы ($v = 0$) получается: для $\varphi = \pi/2$ (доменная граница блоховского типа)

$$\cos\theta = -\tanh\left(\frac{x}{D_{0B}}\right), \ D_{0B} = \sqrt{\frac{\alpha}{\beta}}, \quad (13)$$

а для $\varphi = 0$ (доменная граница неелевского типа)

$$\cos\theta = -\tanh\left(\frac{x}{D_{0N}}\right), \ D_{0N} = x_{0B}\frac{1}{\sqrt{1 + \epsilon}}. \quad (14)$$

При отличной от нуля скорости движения в границе индуцируются дополнительные поля рассеяния, что приводит к возникновению у нее дополнительной энергии. Эта энергия квадратична по скорости: $\frac{1}{2}mv^2$, что дает возможность говорить о коэффициенте $m$ как о массе границы [14]. Дополнительную энергию можно учесть, выполнив перенормировку константы анизотропии и параметра $\epsilon$ [2]

$$\beta \to \beta^* = \beta + 4\pi$$

и

$$\epsilon \to \epsilon^* = \epsilon + 4\pi/\beta,$$

что дает для энергии, приходящейся на единицу площади поверхности границы,

$$\sigma(v) = 2M_0^2\sqrt{\alpha\beta\left(1 + \epsilon^* \cos^2\varphi\right)}. \quad (15)$$

(12) дает также возможность определить предельную (уокеровскую) скорость смещения границы

$$v_W = \frac{1}{2}2D\gamma\beta M_0\left(\sqrt{1 + \epsilon^*} - 1\right). \quad (16)$$

При малых скоростях ($v \ll v_W$)

$$\sigma(v) = \sigma_0 + \frac{m_0 v^2}{2}, \qquad (17)$$

где для блоховской границы

$$m_{0B} = \frac{\sigma_{0B}}{(D\gamma\beta M_0)^2 \epsilon}, \qquad (18)$$

а для неелевской

$$m_{0N} = \frac{-\sigma_{0N}}{(D\gamma\beta M_0)^2 \epsilon (1+\epsilon)^3}. \qquad (19)$$

При приближении скорости границы к пределу Уокера, ее масса и ширина начинают подчиняться квазирелятивистским соотношениям

$$m(v) = \frac{m_0}{\sqrt{1 - v/v_W^2}}, \qquad (20)$$

$$D(v) = D_0 \sqrt{1 - v/v_W^2}, \qquad (21)$$

где $D_0$ — ширина статической доменной границы. В отсутствии затухания уокеровское значение действительно является предельным для скорости границы. Однако, в случае ненулевой диссипации, как было показано Слончевским [18], уокеровский барьер может преодолеваться, но движение границы приобретает неустойчивый характер, а ее подвижность может оказаться отрицательной. Зависимость скорости границы от продвигающего поля дается при этом функцией

$$v(H) \sim 2\pi M_0 \sqrt{\frac{\alpha}{\beta}} \gamma \left( H - \frac{\sqrt{H^2 - 1}}{1 + \lambda^2} \right). \qquad (22)$$

Очень важен с физической точки зрения тот факт, что значение предельной скорости в ферромагнетиках (равно как и в антиферро- и ферримагнетиках) всегда меньше минимальной фазовой скорости спиновых волн. Это обстоятельство делает невозможными резонансные процессы обмена энергией между доменной границей и окружением, так называемые черенковские процессы. Доменные границы в кристалле взаимодействуют с магнитными неоднородностями, создаваемыми либо дефектами кристаллической структуры, либо обусловленными микроскопическим магнитным строением самого кристалла, как это имеет место, например, в

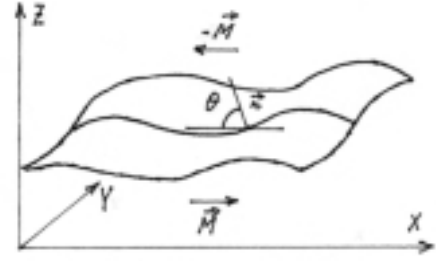

Рис. 2. Доменная граница разделяет области с противоположными направлениями намагниченности $\mathbf{M}$ и $-\mathbf{M}$.

высокоанизотропных магнетиках с собственной коэрцитивностью [20]. Во многих случаях такое взаимодействие удается описывать, вводя эффективный потенциал $U$ взаимодействия границы с неоднородностями. Описание движения границы вдоль оси легкого намагничивания под действием внешнего поля $\mathbf{H}$ сводится тогда к одномерному[4] уравнению вида[5]

$$m\ddot{x} = 2M_0 H - \lambda \dot{x} - \frac{\partial U}{\partial x}, \qquad (23)$$

где $x$ является координатой центра границы.

На протяжении текущей главы, таким образом, мы будем рассматривать доменную границу в качестве квазичастицы с определенной массой $m$, не вдаваясь в детали ее структуры.

Для экспериментального изучения динамики границ применяются разнообразные магнитные, магнитооптические, магнитошумовые и др. методы. Подробный обзор этих методов можно найти, например, в монографии [19]. В рамках нашего рассмотрения мы будем останавливаться на деталях этих методик только в тех случаях, когда от деталей методов измерения зависят физические особенности изучаемых процессов. В общем случае достаточно полагать, что быстрота изменения суммарного магнитного момента кристалла при смещении одной границы будет определяться как $\dot M = 2M_0 vS$, где $S$ — площадь смещающегося участка границы. Таким образом, отслеживая изменение намагниченности в различных экспериментальных ситуациях, можно получить достаточно подробные сведения о характере динамики границы.

---

[4]Учет внутренних степеней свободы, вообще говоря, увеличивает размерность уравнения. Детальное обсуждение этой ситуации мы проводим в главе 5.

[5]Все величины приводятся в расчете на единицу площади.

## 1.2 Динамика доменных границ в антиферромагнетиках

В соответствии с [22] магнитная структура антиферромагнетика может быть представлена в двухподрешеточной модели локальными плотностями магнитных моментов $\mathbf{M}_1(\mathbf{r},t)$ и $\mathbf{M}_2(\mathbf{r},t)$. Векторы антиферромагнетизма $\mathbf{L}$ и намагниченности $\mathbf{M}$ определяются как

$$\mathbf{L} = \mathbf{M}_1 - \mathbf{M}_2, \ \mathbf{M} = \mathbf{M}_1 + \mathbf{M}_2.$$

причем
$$\mathbf{ML} = 0, \ \mathbf{M}^2 + \mathbf{L}^2 = 4M_0^2.$$

Для одноосного антиферромагнетика магнитная энергия дается выражением

$$W = \frac{1}{2}A\mathbf{M}^2 + \frac{1}{2}\alpha_1 \left(\frac{\partial \mathbf{L}}{\partial x_i}\right)^2 + \frac{1}{2}\alpha_2 \left(\frac{\partial \mathbf{M}}{\partial x_i}\right)^2 - \frac{1}{2}\beta_1 L_z^2 - \frac{1}{2}\beta_2 M_z^2 - \mathbf{MH} \quad (24)$$

где $A$ — константа однородного обмена, $\alpha_1$, $\alpha_2$ — константы неоднородного обмена, $\beta_1$, $\beta_2$ — константы одноосной анизотропии.

При условии малости энергии релятивистских взаимодействий по сравнению с энергией обмена уравнения для $\mathbf{M}$ и $\mathbf{L}$ могут быть сведены к уравнениям только для $\mathbf{L}$ [23]. Такое эффективное уравнение имеет вид

$$\left[\mathbf{L}\left(c^2\Delta\mathbf{L} - \frac{\partial^2 \mathbf{L}}{\partial t^2}\right)\right] =$$
$$= \frac{4\mu_0}{\hbar}(\mathbf{LH})\frac{\partial \mathbf{L}}{\partial t} + \frac{4\mu_B^2}{\hbar^2}(\mathbf{LH})[\mathbf{L}\times\mathbf{H}] - \omega_0^2 L_z [\mathbf{L}\times\mathbf{e}_z], \quad (25)$$

где
$$c = \frac{4\mu_B M_0}{\hbar}\sqrt{A\alpha_1},$$
$$\omega_0 = \frac{4\mu_0}{\hbar}\sqrt{A\beta_1}.$$

Вводя вектора
$$\mathbf{m} = \frac{\mathbf{M}}{2M_0}, \ \mathbf{l} = \frac{\mathbf{L}}{2M_0},$$

и выбирая в качестве единицы длины $c/\omega_0$, и в качестве единицы времени $1/\omega_0$, уравнение (25) можно записать в безразмерном виде

$$\mathbf{l}\left(\Delta\mathbf{l} - \frac{\partial^2 \mathbf{l}}{\partial t^2}\right) - 2\frac{\partial \mathbf{l}}{\partial t}(\mathbf{lh}) + (1-h^2)[\mathbf{l}\times\mathbf{e}_z](\mathbf{le}_z) = 0,$$
(26)

где $\mathbf{h} = \mathbf{H}/2\sqrt{A\beta_1}$. Вводя переменные $l_x = \sin\theta\cos\varphi$, $l_y = \sin\theta\sin\varphi$, $\theta$, (26) можно записать в виде

$$c^2\frac{\partial^2 \theta}{\partial \xi^2} - \frac{\partial^2 \theta}{\partial t^2} +$$
$$+ \left[\left(gH - \frac{\partial \varphi}{\partial t}\right)^2 - c^2\left(\frac{\partial \varphi}{\partial \xi}\right)^2 - \omega_0^2\right]\sin\theta\cos\theta = 0$$
(27)

$$c^2\frac{\partial}{\partial \xi}\left(\sin^2\theta \frac{\partial \varphi}{\partial \xi}\right) = \frac{\partial}{\partial t}\left[\sin^2\theta\left(-gH + \frac{\partial \varphi}{\partial t}\right)\right].$$

В антиферромагнетике энергия доменной границы в расчете на единицу площади будет [23]

$$\sigma(v) = 8M_0^2\sqrt{\alpha_1\beta_1}\frac{1 - \left(H/2\sqrt{A\beta_1}\right)^2}{\sqrt{1 - \left(H/2\sqrt{A\beta_1}\right)^2 - (v/c)^2}}.$$
(28)

При малых скоростях ($v \ll c$), подобно (16), энергия границы может быть представлена как

$$\sigma(v) = \sigma_0 + m_0 v^2/2,$$

где $\sigma_0 = 8M_0^2\sqrt{\alpha_1\beta_1(1 - H^2/4A\beta_1)}$, и $m_0 = \dfrac{\hbar^2\beta_1}{\mu_B^2\sqrt{A\alpha_1(4A\beta_1 - H^2)}}$. Ясно, что и в этом случае остаются в силе квазирелятивистские соотношения, справедливые для границ в ферромагнетиках.

## 1.3 Доменные границы в слабых ферромагнетиках

Динамика границ в слабых ферромагнетиках, в определенном смысле, демонстрирует гораздо большее разнообразие новых физических явлений в сравнении с ситуацией в ферро- и антиферромагнетиках. Это сделало их крайне привлекательными объектами для экспериментального и теоретического изучения. Например, тот факт, что предельная скорость движения границ в слабых ферромагнетиках выше скорости звука, дает возможность реализоваться очень специфическим процессам взаимодействия границ с упругими и тепловыми колебаниями кристалла. Не менее интересные особенности возникают и в протекании существенно квантовых явлений, таких как туннелирование. Кроме того, слабые ферромагнетики, в силу совершенства их кристаллической структуры являются очень удобными объектами для экспериментального изучения.

Разработка теории динамики намагниченности в слабых ферромагнетиках началась в конце 70-х годов [21], [25]. Следуя [21] и [24] получим динамические уравнения для границы в слабом ферромагнетике. Будем рассматривать слабый ферромагнетик ромбической симметрии. Он может быть описан в двухподрешеточном приближении с помощью векторов ферромагнетизма $\mathbf{m}$ и антиферромагнетизма $\mathbf{l}$. Термодинамический потенциал $\Phi(\mathbf{l}, \mathbf{m})$ дается выражением

$$\Phi(\mathbf{l}, \mathbf{m}) = Jm^2 + A(\nabla \mathbf{l})^2 - \mathbf{mH} + d_1 m_x l_z - \\ - d_3 m_z l_x + \beta_{ac} l_z^2 + \beta_{ab} l_x^2. \quad (29)$$

Здесь $J$ — константа однородного обмена, $A$ — константа неоднородного обмена, $\beta_{ac}$ и $\beta_{ab}$ — константы анизотропии, $\mathbf{H}$ — внешнее поле, $d_1, d_3$ — постоянные антисимметричного обмена Дзялошинского, $\mathbf{m}$ и $\mathbf{l}$ связаны соотношениями

$$\mathbf{lm} = 0, \quad l^2 = 1 - m^2 \approx 1.$$

После минимизации (29) по $\mathbf{m}$, $\Phi$ получается в виде

$$\Phi = A(\nabla \mathbf{l})^2 - \frac{\chi_\perp}{2}\left(H^2 - (\mathbf{Hl})^2\right) - M_z^0 H_z l_x - \\ - M_x^0 H_x l_x + \beta_{ac} l_z^2 - \beta_{ab} l_x^2,$$

где $\chi_\perp = \frac{M_0}{2H_E}$ — поперечная восприимчивость, $M_z^0, M_x^0$ — компоненты вектора ферромагнетизма в фазах $\Gamma_4 (G_x A_y F_z)$ и $\Gamma_2 (F_x C_y G_z)$.

Для доменной границы $ac$−типа в сферической системе координат $l_x = \sin\theta\cos\varphi$, $l_y = \sin\theta\sin\varphi$, $l_z = \cos\theta$, объемная плотность лагранжиана будет [26]

$$\mathcal{L} = \frac{\chi_\perp}{2\gamma^2}\left(\dot{\mathbf{l}}\right)^2 - \frac{\chi_\perp}{\gamma}\mathbf{H}\left[\mathbf{l} \times \dot{\mathbf{l}}\right] - \Phi. \quad (30)$$

В сферической системе координат для границы $ac$−типа, движущейся в отсутствии внешнего поля со скоростью $v$, $\varphi = 0$, решение же для $\theta$ получается следующим образом. После подстановки $\xi = \frac{x - vt}{D}$, где $D = D_0\sqrt{1 - \frac{v^2}{c^2}}$, а $D_0 = \sqrt{A/(\beta_{ac} + \chi_\perp H^2)}$ — ширина покоящейся доменой границы и $c = \gamma\sqrt{A/\chi_\perp}$ — предельная скорость доменной границы, совпадающая со скоростью спиновых волн, (30) принимает вид

$$\frac{\partial^2 \theta}{\partial \xi^2} = -\sin\theta\cos\theta. \quad (31)$$

Выбором решения (31) в виде

$$\theta_0 = -\frac{\pi}{2} + 2\arctan e^\xi$$

осуществляется переход к сокращенному описанию динамики границы. В таком подходе доменная граница рассматривается в качестве частицы, координата которой $x_0(t)$ совпадает с координатой центра границы и связана с переменной $\theta$ соотношением

$$\theta(t) = \theta_0\left(\frac{x - x_0(t)}{D(t)}\right),$$

где $D(t) = D_0\sqrt{1 - \left(\frac{\dot{x}_0(t)}{c}\right)^2}$. Выбирая поле направленным вдоль оси $c$ и интегрируя плотность лагранжиана в пределах одного домена, можно получить для функции Лагранжа следующее выражение[6]

$$L = -mc^2/\sqrt{1 - v^2/c^2} - U(x_0) \qquad (32)$$

где $v = \dot{x}_0$, $mc^2 = 2\sqrt{A\left(\beta_{ac} + \chi_\perp H^2\right)}$, а эффективный потенциал есть $U(x_0) = -\int 2M_z^0 H(x_0)\, dx_0$ с $H(x_0)$, и представляет эффективное поле, включающее внешнее поле и поле дефекта.

Лагранжиану (32) соответствует функция Гамильтона

$$H(p, x) = c\sqrt{p^2 + m^2 c^2} + U(x), \qquad (33)$$

с каноническим импульсом

$$p = \frac{mv}{\sqrt{1 - v^2/c^2}}. \qquad (34)$$

## 1.4 Одномерная динамика доменных границ. Взаимодействие с магнитными неоднородностями

Несмотря на простой вид уравнения (23) динамика границ магнитных доменов даже в одномерной модели[7] является далеко не тривиальной. Это обусловлено следующими факторами:

— потенциальный рельеф $U(x)$, формируемый магнитными дефектами кристалла вместе с внешним полем, чаще всего имеет достаточно сложную форму, не позволяющую получить решение в аналитическом виде;

— коэффициент $\lambda$ может оказаться достаточно сложной функцией скорости границы, что связано с резонансным характером взаимодействия границы с окружением;

— доменные границы являются мезоскопическими объектами, что требует обязательного учета квантовых эффектов, например, туннелирования.

Начнем с обсуждения первого фактора. Причины формирующие магнитные неоднородности могут быть достаточно разнообразны. Такие неоднородности могут создаваться точечными дефектами типа атомарных примесей или вакансий, протяженными дефектами, например, дислокациями или дисклинациями. Магнитные неоднородности могут быть модулированы объемными или поверхностными неоднородностями образца.

Информация о характере неоднородностей и их распределении в образце (определяющая, в конечном счете, вид $U(x)$) может быть получена лишь косвенно по данным о смещении границы вдоль кристалла под действием приложенного поля. Методы получения такой информации также могут быть различны. Это могут быть данные прямых магнитооптических наблюдений за смещением границы, гистерезисные измерения, магнитошумовые методы, данные об изменении электрического сопротивления, сопровождающего смещение границ [19].

Однако даже в случае наблюдения за смещением уединенной границы информация о пространственном распределении потенциальной энергии имеет существенно стохастический характер. Речь идет, конечно, о том, что полная энергия границы не является детерминированной — в нее входит стохастический вклад, обусловленный тепловым взаимодействием с кристаллом — и может быть определена лишь усредненно

$$\langle W \rangle = W_H + \langle W_T \rangle$$

(здесь $W_H$ — энергия взаимодействия границы с магнитным полем, $\langle W_T \rangle$ — тепловая энергия границы).

Таким образом, локальная амплитуда рельефа $U_0$, вычисленная по экспериментальному значению поля срыва $H_c$ не может рассматриваться как однозначно определенная величина. Это обстоятельство требует достаточно аккуратно относиться к экспериментальным измерениям параметров рельефа, не подтвержденным достаточной статистикой. По этой же причине очень чувствительными могут оказаться вычисления параметров рельефа и к временным характеристикам измерений.

Для иллюстрации методов вычисления потенциального рельефа $U(x)$ рассмотрим простейшую ситуацию взаимодействия границы с точечной неоднородностью в квазиодномерном ферромагнитном кристалле. В силу того, что граница имеет конечную ширину, такое взаимодействие оказывается нелокальным и не описывается $\delta$-функцией. В соответствии с (9) полная энергия ферромагнетика в одномерном случае может быть представлена как

$$W = S \int_0^L dx \{ A \left[ \left( \frac{\partial \theta}{\partial x} \right)^2 + \sin^2\theta \left( \frac{\partial \varphi}{\partial x} \right)^2 \right] +$$

$$+ K_e \sin^2\theta + K_H \sin^2\theta \sin^2\varpi \}, \qquad (35)$$

где введены обозначения $L$ — длина образца, $S$ — его площадь поперечного сечения, $A = \alpha M_0^2/2$, $K_e$ и $K_h$ — перенормированные константы анизотропии соответственно вдоль легкой и трудной осей.

---
[6]В расчете на единицу площади границы.

[7]Пространственная ось $x$ совпадает с направлением легкой оси кристалла **e**, вдоль которой приложено внешнее поле.

Пусть неоднородность создана атомом, обусловливающим дополнительную анизотропию вдоль легкой оси в точке $x = 0$ $K_d \neq K_e$. Переобозначим

$$K_e \to K_e + K_d(x), \; K_d(x) = -U_0 \delta(x), \qquad (36)$$

где $U_0 = (K_e - K_d)a^3$, $a$ — параметр решетки кристалла. Для притягивающей примеси $U_0 > 0$, для отталкивающей — $U_0 < 0$. Дополнительная энергия, связанная со взаимодействием с дефектом и с внешним полем дается выражением

$$W' = \int dx \left[ K_d(x) - H M_0 \cos\theta \right].$$

Ограничимся случаем слабого потенциала, т.е. такого, для которого энергия $U_0$ много меньше статической энергии границы $W_0 = 2N K_d a^3$ [8]. В этом случае изменением формы границы можно пренебречь. Подставляя далее в выражение для $W'$ функцию $\theta(x) = arcsin(\tanh(x/\sqrt{AK_e}))$ получаем

$$W'(x) = U_d(x) - 2M_0 H x, \qquad (37)$$

где

$$U_d(x) = -\frac{U_0}{\cosh^2\left(\frac{x}{D}\right)} \qquad (38)$$

есть искомый потенциал взаимодействия границы с точечной неоднородностью.

Отметим, что потенциал вида (38) получается не только для случая взаимодействия с точечной

---

[8] $N = 2SD/a^3$ — число спинов в статической доменной границе.

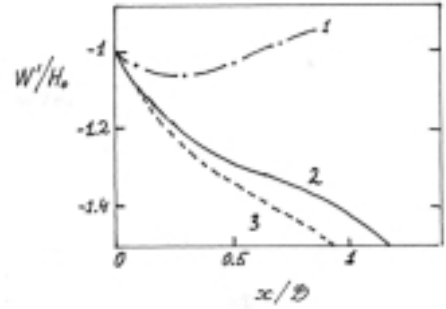

Рис. 3. Вид функции (37) для трех различных значений продвигающего поля: 1 - $H/H_c = 0.5$, 2 - $H/H_c = 1$, 3 - $H/H_c = 1.1$, где $H_c$ - коэрцитивная сила. Видно, что потенциальный барьер для границы полностью исчезает при значениях $H$, превышающих $H_c$.

неоднородностью. В точности такую же форму будет иметь потенциал взаимодействия границы, распространяющейся в тонком ферромагнитном образце с модулированной толщиной, когда ширина участка с измененной толщиной много меньше ширины границы. Действительно, пусть в процессе своего движения граница попадает в область образца, где площадь поперечного сечения локально уменьшается $S(x) = S - \Delta S(x)$. Это приводит к локальному уменьшению объема на величину $\Delta V = \int dx \Delta S(x)$. Тогда $W'$ будет

$$W' = -2K_e \Delta V \operatorname{sech}^2\frac{x}{D} - 2M_0 H x,$$

а $U_0 = 2K_e \Delta V$.

Потенциальный рельеф для границы, движущейся в образце, содержащем много дефектов, в принципе, может быть рассчитан суперпозицией выражений (38). Однако, реально экспериментальная задача получения детальной информации о распределении и величине магнитных неоднородностей в многодефектном образце является очень сложной, и до настоящего времени не имеет удовлетворительного решения. Речь может идти лишь об определении усредненных интегральных характеристик распределения дефектов в образце, таких как их концентрация, или среднее распределение энергий магнитных неоднородностей. Поэтому здесь мы ограничимся рассмотрением взаимодействия границы с единичными дефектами. На самом деле, успехи экспериментальной техники позволили решить задачу получения образцов с чрезвычайно низкой концентрацией дефектов (см., например, [27]), так что их изучение представляется вполне актуальным.

## 1.5 Влияние термоактивации и туннелирования на макроскопические характеристики процессов перемагничивания

Доменная граница, рассматриваемая как классическая частица, попадая в область потенциального барьера, созданного дефектом, в зависимости от соотношения между ее энергией во внешнем поле $W$ и высотой барьера $U_0$ либо замедляет скорость, но преодолевает дефект в случае $W > U_0$, либо замедляется и останавливается в случае $W \leq U_0$. Однако, при таком подходе не учитывается, что даже в случае $W \leq U_0$ граница может получить дополнительную энергию за счет термоактивации или преодолеть барьер за счет туннелирования.

В этом параграфе мы продемонстрируем, как учет термоактивации или туннелирования, сказывается на макроскопических характеристиках процессов перемагничивания, таких, например, как коэрцитивная сила.

Пусть в момент времени $t = 0$ граница находилась в потенциале $U(x)$ в точке $x = 0$ перед барьером высотой $U_0$ В это время включается монотонно (для определенности — линейно) нарастающее внешнее поле - $H(t) = at$. К моменту времени $t$ поле сообщит границе энергию $W(t) = U(x(t))$, где $x(t)$ определяется решением уравнения

$$-\frac{\partial U}{\partial x} = 2M_0 at. \qquad (39)$$

Например, для квадратичного потенциального барьера

$$U(x) = -k(x - a)^2 + U_0$$

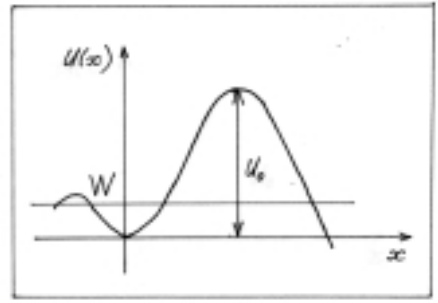

Рис. 4. Энергия границы во внешнем поле может быть определена как $W = U(x)$, где $x$ дается решением уравнения (39).

$x(t)$ дается выражением $x = \frac{M_0 t}{k}$ и

$$W(t) = -k(\frac{M_0 t}{k} - a)^2 + U_0.$$

Таким образом, энергия границы в любой момент времени может быть определена, как сумма энергии $W(t)$ и тепловой энергии $W_T$. Условие преодоления границей препятствия $W(t) + W_T \geq 0$, поэтому вероятность срыва в момент времени $t$ будет[9]

$$P_{therm}(t, T) =$$
$$= \int_{U_0 - W(t) - W_T}^{\infty} \frac{2}{\sqrt{\pi}} (k_B T)^{-3/2} w^{1/2} \exp\left(-\frac{w}{k_B T}\right) dw. \qquad (40)$$

Мы используем здесь при анализе срыва границ распределение Максвелла. Как будет показано ниже, чаще всего в ферромагнетиках параметры ширины границы, эффективной массы, высоты и ширины барьера таковы, что энергетический спектр границы можно считать квазинепрерывным, что делает допустимым подобное квазиклассическое рассмотрение.

---

[9]Скорости поля в экспериментах составляют обычно $0.01 - 1$ T/s, обычные же времена релаксации в ферромагнетиках $\tau_0$ составляют $10^{-4} - 10^{-2}$ s. Таким образом, границу в реальных измерениях можно считать находящейся в тепловом равновесии с окружением.

записать
$$M(t,T) = M_0 N S x(\bar{t}, T), \qquad (42)$$
где $S$ — площадь поперечного сечения образца, $N$ - число доменных границ в образце.

Заметим, что в выражении (42) $P(t,T)$ — это просто вероятность срыва, безотносительно к тому, какими причинами этот срыв вызван — термоактивацией, туннелированием, или совместным действием названных механизмов. Поэтому в дальнейших расчетах кривых намагничивания, связанных с исследованием других механизмов срыва, мы будем пользоваться этим же выражением. Учет неоднородности характеристик дефектов, в принципе, можно произвести по описанной выше схеме. Однако, поскольку на данном этапе мы ставим перед собой задачу исследования именно механизмов, приводящих к срыву, а не расчета кривых намагничивания, детального обсуждения по этому поводу здесь проводиться не будет.

В некоторых случаях бывает полезно вычисление эффективного снижения коэрцитивной силы в результате протекания термоактивационных или туннельных процессов срыва. Будем называть коэрцитивной силой такое значение продвигающего поля $H_c$, при котором справедливо равенство

$$2M_0 H_c S = -\left(\frac{\partial U}{\partial x}\right)_{x=x_{max}}. \qquad (43)$$

Например для потенциала, созданного *отталкивающей* точечной примесью, вида аналогичного потенциалу (38)

$$U(x) = U_0 \cosh^{-2}\left(\frac{x}{D}\right),$$

получаем

$$F(x) = 2\frac{U_0}{\cosh^3 \frac{x}{D}} \frac{\sinh \frac{x}{D}}{D}.$$

Максимум для $F(x)$ достигается в точке

$$x = \left(arccosh\left(\sqrt{3/2}\right)\right)D,$$

где сила равна

$$F_{\max} = 0.77 \frac{U_0}{D}.$$

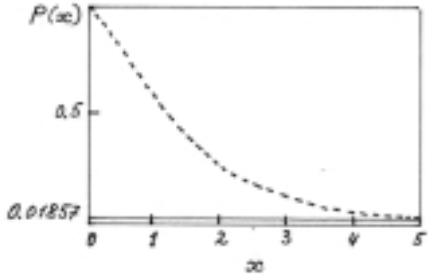

Рис. 5. Вид функции (41)

Интегрирование (40) дает следующий вид зависимости вероятности срыва от температуры при фиксированном значении $t$

$$P(\eta) = 1 + \frac{2}{\sqrt{\pi}} e^{-\eta} \sqrt{\eta} - erf(\sqrt{\eta}), \qquad (41)$$

где $\eta = \frac{U_0 - W(t) - W_T}{k_B T}$.

Вид функции (41) мы приводим на рис. 1.5, ясно, что на самом деле $P$ является функцией времени и температуры. Несложно убедиться, что зная вид $P(t,T)$, можно восстановить вид кривой намагничивания, как функции времени. Продемонстрируем это для квазиодномерного образца, содержащего ограниченное число доменных границ. Вначале рассмотрим случай, когда все дефекты в образце идентичны[10], однако распределены вдоль образца случайным образом.

Пусть $\bar{\lambda}$ — среднее расстояние между дефектами, причем $\bar{\lambda} \gg D$, т.е. порядка ширины потенциального барьера, совпадающей для точечных дефектов с шириной доменной границы. При движении границы намагниченность будет увеличиваться пропорционально ее смещению. Среднее же смещение доменной границы, в свою очередь, дается выражением $x(\bar{t},T) = \int_0^t \bar{\lambda} P(\tau,T) d\tau$. Если пренебречь квазистатическим смещением границ в пределах области влияния дефекта (шириной $D$), то для намагниченности, как функции времени и температуры, можно

---

[10]Т.е. создают потенциальные барьеры одинаковой высоты $U_0$ и одинаковой формы

Соответствующая коэрцитивная сила будет

$$H_c = \frac{0.77 U_0}{2 M_0 SD}. \qquad (44)$$

Ясно, что при любой ненулевой температуре значение коэрцитивной силы, определяемое (44) достигаться не будет — срыв наступит раньше в поле

$$H_c = \frac{0.77 \left(U_0 - \langle W_T \rangle\right)}{2 M_0 SD}, \qquad (45)$$

где $\langle W_T \rangle$ — средняя энергия термоактивации [11].

Менее очевидно, как скажется влияние туннелирования на величине коэрцитивной силы. Будем полагать, что доменная граница находится под воздействием некоторого поля $H < H_c$, приложенного вдоль оси легкого намагничивания, перед дефектом, создающим потенциальный барьер высотой $U_0$.

Пусть вероятность преодоления барьера границей путем туннелирования дается величиной $P_{tun}(H, U_0)$. С этой вероятностью можно формально связать некоторую дополнительную энергию $W_{tun}$, эквивалентную величине тепловой энергии, при которой будет возникать термоактивационный срыв с точно такой же по величине вероятностью $P_{therm}$, определяемой выражением (40), т.е. будем полагать $W_{tun} \longleftrightarrow W_T$ и $P_{tun} \longleftrightarrow P_{therm}$. В силу (41) можно написать теперь

$$P_{tun} = 1 + \frac{2}{\sqrt{\pi}} e^{-\eta} \sqrt{\eta} - erf(\sqrt{\eta}), \qquad (46)$$

где $\eta$ теперь вместо $W_T$ содержит $W_{tun}$:

$$\eta = \frac{U_0 - W(t) - W_{tun}}{k_B T}.$$

(46) является неявным уравнением для определения эффективной энергии $W_{tun}$. Так как вероятность туннелирования приобретает сколь-нибудь заметное значение лишь для малых $\eta$, в (46) последним членом можно пренебречь. В этом случае $\eta$ можно записать в виде

$$\eta = \frac{1}{4} \pi \exp^2 \left( -\frac{1}{2} LambertW \left( -\frac{1}{2} \pi \left(-1 + P_{tun}\right)^2 \right) \right) \times \left(-1 + P_{tun}\right)^2,$$

---
[11] Мы по-прежнему предполагаем процесс перемагничивания адиабатическим.

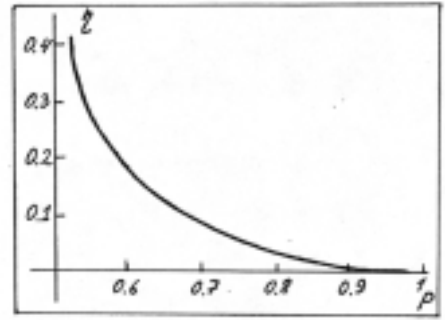

Рис. 6. Зависимость $\eta(P_{tun})$.

где $LambertW(x)$ — функция Ламберта, определяемая уравнением

$$LambertW(x) e^{LambertW(x)} = x.$$

Таким образом, для малых $\eta$ явное выражение для дополнительной «туннельной» энергии имеет вид

$$W_{tun} = U_0 + \\ + \frac{1}{2} LambertW \left( -\frac{1}{2} \pi \left(-1 + P_{tun}\right)^2 \right) k_B T - W_t. \qquad (47)$$

Выражение для коэрцитивной силы, эффективно снизившей свое значение из-за наличия туннелирования полностью аналогично (45), с заменой $W_T$ на $W_{tun}$.

Таким образом, при вычислениях высоты потенциальных барьеров дефектов с использованием полученных экспериментально значений коэрцитивной силы нужно обязательно иметь в виду, что в реальных измерениях значение коэрцитивной силы всегда получается заниженным в сравнении с ее истинным значением, определяемым (43). Это обстоятельство обусловлено эффектами термоактивации и туннелирования.

В дальнейшем, при сравнении расчетных результатов, полученных в настоящей работе, с данными экспериментов, мы будем пользоваться для оценок высоты барьеров выражениями вида (45), в которых, учтена возможность эффективного занижения высоты барьера.

Таблица 1. Намагниченности насыщения $M_0$ и константы анизотропии вдоль трудной $\beta_h$ и легкой $\beta_e$ осей для разных типов магнетиков.

|      | $M_0$ $(Oe)$ | $\beta_h$ $(10^5 \frac{erg}{cm^3})$ | $\beta_e$ $(10^5 \frac{erg}{cm^3})$ |
|------|------|------|------|
| $YIG$ | 196 | 1.4 | 0.25 |
| $Ni$  | 508 | 16  | 8 |
| $REC$ | 200 | $3-5$ | $100-500$ |

Таблица 2. Постоянная обмена $A$, ширина границы $D$, деринговская масса $m_D$ и коэрцитивная сила $H_c$ для различных типов магнетиков.

|      | $A$ $10^{-6}\frac{erg}{cm}$ | $D$ Å | $m_D$ $(10^{-10}\frac{g}{cm^2})$ | $H_c$ $(Oe)$ |
|------|------|------|------|------|
| $YIG$ | 0.43 | 414 | 1.2 | 10 |
| $Ni$  | 1 | 112 | 4.6 | 100 |
| $REC$ | 1 | $5-30$ | 10 | 10 |

В справочных целях, в таблицах 1.1 и 1.2 мы приводим значения основных динамических параметров для границ в магнетиках различных типов (цитируется по [19] ).

# 2 Туннелирование границ магнитных доменов в ферромагнетиках

Несмотря на то, что первые теоретические предсказания относительно возможности наблюдения эффекта туннелирования границ магнитных доменов были сделаны в начале 70-х годов [6], [20], [29], первые эксперименты, в которых наблюдались особенности, допускающие интерпретацию в качестве проявления туннелирования, стали выполняться лишь в середине 90-х годов [27], [30], [31], [32], [33].

Практически во всех экспериментальных работах основным признаком для обнаружения туннелирования стало изменение характера магнитной релаксации, как правило, в области гелиевых и субгелиевых температур. Ниже мы более подробно обсудим некоторые эксперименты, в которых подобные эффекты наблюдались. Однако, в большинстве случаев, даже для систем, заведомо допускающих описание в рамках одномерных моделей [27], наблюдаются определенные сложности в согласовании теоретических и экспериментальных результатов. Простые расчеты приводят для конкретных экспериментальных ситуаций к результатам, которые означают вообще полную невозможность туннелирования.

Поэтому в данном разделе мы уделим основное внимание факторам, которые как раз могли бы обеспечить увеличение вероятности туннелирования. К ним мы будем относить:

— взаимодействие границы с тепловой системой кристалла, приводящее к эффективному снижению высоты барьера [34];

— квазирелятивистские эффекты, приводящие к эффективному снижению ширины барьера [35];

— эффекты резонансного поглощения энергии границей в процессе преодоления потенциального барьера [44], [36].

Раздел имеет следующую структуру. В начале, с целью выбора расчетных моделей и определения количественных параметров теории, мы приводим обзор известных из литературы экспериментальных результатов. В следующих параграфах анализируются вычислительные методы, используемые для описания туннелирования доменных границ. Далее проводится обсуждение факторов, способствующих увеличению вероятности туннелирования.

## 2.1 Туннелирование границ в ферромагнетиках. Эксперименты

Уверенной идентификации туннельных процессов срыва доменных границ в ферромагнетиках препятствуют вполне очевидные обстоятельства: 1) В большинстве случаев доменная структура как массивных, так и пленочных ферромагнитных образцов, является достаточно сложной; применение к ее описанию методов понижения размерности задачи, изложенных в предыдущем разделе (например, «сокращенного» описания [25], к сожалению, затруднено. 2) Как правило, в реальных образцах ферромагнитных материалов существует огромное разнообразие дефектов, различающихся своими размерами и формой. При этом нужно учитывать, что и распределение границ, и распределение дефектов носят в достаточной мере случайный характер. Выделить на фоне этих двух факторов стохастизации такой также существенно стохастический процесс, как туннелирование, тем более в присутствии термоактивации, представляет собой действительно сложную задачу.

Поэтому основные усилия экспериментаторов в этой области были направлены на получение образцов с хорошо предсказуемой доменной структурой и возможно более достоверно известным распределением потенциального рельефа для движущихся границ. Такие объекты удалось получить в середине 90-х годов. Ими стали сверхтонкие[12] проволоки, изготовленные из ферромагнитных материалов - $Ni$ в [27], [31], и $Fe$ в [32].

Получаются такие проволоки методом испарения тонких ферромагнитных пленок в вакууме, и их толщина составляет обычно несколько сот ангстрем. Преимущества таких образцов очевидны. Во-первых, их можно рассматривать как квазиодномерные объекты, со всеми вытекающими пре-

---
[12]В некоторых источниках их называют мезоскопическими.

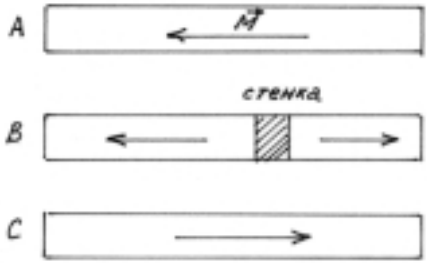

Рис. 7. Доменная структура в мезоскопической проволоке $Ni$. (Взято из [31]).

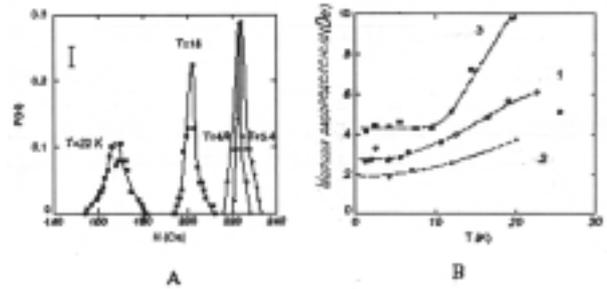

Рис. 8. Экспериментальные данные [27] . **A** — распределение полей срыва при нескольких температурах, **B** — ширина распределения полей срыва как функция температуры для трех различных образцов.

имуществами при выборе математической модели. Во-вторых, в этих образцах можно достаточно надежно контролировать присутствие дефектов, точнее говоря, образцы, в принципе могут получаться идеально чистыми, а магнитные неоднородности создаются искусственно, например, путем модуляции толщины проволоки. В третьих, в таких объектах достаточно легко создавать требуемую доменную структуру, в частности, в цитированных выше экспериментах измерения выполнялись для образцов с уединенными доменными границами. В качестве зонда процессов релаксации для таких объектов использовалось удельное электросопротивление материала. Последнее очень чувствительно к намагниченности образца [37] и, следовательно может использоваться для наблюдения за динамикой доменных границ. В качестве доказательства присутствия туннелирования в процессах релаксации в таких образцах в [27] приводится факт сужения распределения полей срыва по мере понижения температуры.

При более высоких температурах, где срыв обусловлен термоактивацией стохастизация выше, чем при более низких температурах, где срыв предполагается обусловленным туннелированием (рис. 1.8 A). Дополнительным аргументом в пользу туннельной версии служит и то обстоятельство, что, начиная с некоторых значений, дальнейшее понижение температуры уже не сказывается на ширине распределения (рис. 1.8 B).

Данные, аналогичные данным [27] были получены в [32], но уже для железных проволок.

Приведем некоторые оценки параметров задачи, которые несложно получить, используя данные [27]. Диаметр никелевой проволоки примем величиной $4 \cdot 10^{-6} cm$. Для ширины доменной границы $D$ в [27] приводится значение $10^{-5} cm$. Предполагая для константы обмена $A$ обычное значение $1 \cdot 10^{-6} erg/cm$, и учитывая приведенное выше значение $D$, для константы анизотропии получаем $\beta = 5 \cdot 10^{-3} erg/cm^3$. Площадь туннелирующего участка можно оценить по данным рис. 1.8 с помощью очевидного соотношения

$$S \approx \frac{k_B(T_2 - T_1)}{2M_0(H_{r2} - H_{r1})a},$$

где $H_{r2}$ и $H_{r1}$ — правые границы распределения полей старта при температурах $T_2$ и $T_1$ соответственно. Это дает для $S$ порядок $10^{-14} cm^2$. Поверхностная плотность энергии доменной границы будет $\sigma = 2\sqrt{A\beta}$, а ее деринговская масса определяется выражением $m_D = \sigma/(8\pi A\gamma^2)$. Для обсуждаемых материалов деринговская масса равна $m_D = 1.7 \cdot 10^{-11} \ g/cm^2$, а эффективная масса туннелирующего участка составляет величину $m = m_D S \sim 10^{-25} \ g$. С помощью того же рисунка 2.8 можно оценить истинное поле старта[13] $H_c = 290 \, Oe$, что дает для высоты барьера значение $U_0 S = 1.5 \cdot 10^{-14} \, erg$, наблюдаемое же значение $H_c$ составляет около $220 \, Oe$. Ясно, что при таких параметрах прозрачность оценивается действительно исчезающе малыми величинами[14], во всяком случае, она не выше $10^{-70}$.

---

[13] Т.е. коэрцитивную силу в смысле определения (43).

Здесь нужно отметить, что в части, касающейся целенаправленного поиска туннелирования границ в ферромагнетиках, насколько нам известно, помимо экспериментов [27], [31], [32] к моменту написания настоящей работы не было известно о подобных измерениях, выполненных другими авторами. В то же время существует обширный экспериментальный материал, который дает возможность по косвенным признакам предположить, что в тех или иных ситуациях имеет место туннелирование границ. Подробный обзор по этому поводу можно найти, например, в [39].

## 2.2 Туннелирование границ в слабых ферромагнетиках

В отличие от ситуации с экспериментальным наблюдением туннелирования границ в ферромагнитных мезоскопических образцах, где в качестве критерия идентификация туннелирования служило аномальное поведение электропроводности, основным критерием для распознавания туннелирования в экспериментах со слабыми ферромагнетиками служит отклонение скорости процессов магнитной релаксации от закона Аррениуса. Такие аномалии описаны в [40] для тербиевого ортоферрита. Оценки основных динамических характеристик для границ в этом материале были получены в [24]. Если принять, что взаимодействие границы с дефектом моделируется полем

$$H(x) = H_c \left(1 - \frac{(x-b)^2}{2a^2}\right),$$

где $H_c$ — коэрцитивная сила, $a$ — ширина дефекта, $b$ — координата его центра, то для эффективного потенциала барьера, с учетом воздействия внешнего поля $H_0$, можно записать

$$U(x) = \frac{M_z^0 H_c}{3a^2} x^3 + \frac{M_z^0 H_c}{a} x^2 \sqrt{2\left(1 + \frac{H_0}{H_c}\right)}.$$

Согласно [40] $\beta_{ac} = 1 \cdot 10^5 \, erg/cm^3$, постоянная обмена $A = 1 \cdot 10^{-7} \, erg/cm$, $M_z^0 = 10 \, Gs$, предельная скорость границы $c = 2 \cdot 10^6 \, cm/s$. Это дает для ширины границы $D = 10^{-6} \, cm$, для поверхностной плотности энергии $\sigma_0 = mc^2 = 4\sqrt{A\beta_{ac}} = 0.4 \, erg/cm^2$, и для высоты барьера $U_{max}^0 = mc^2 \frac{d}{2D}$, где $d = 10 \text{Å}$ — длина дефекта. Площадь туннелирующего участка в [24] оценивается как $S \approx 4 \cdot 10^3 \, \text{Å}^2$, а его объем $V = \frac{U}{2M_z^0 H_c} \approx 7 \cdot 10^5 \, \text{Å}^3$. Ширина барьера принимается равной ширине границы, т.е. дефект считается практически точечным.

При приведенных выше значениях параметров показатель туннельной экспоненты получается приблизительно равным (-30). Иначе говоря, также как и в случае, описанном в [27], теория предсказывает существенно более низкую вероятность туннелирования в сравнении с наблюдающейся в экспериментах.

Совсем недавно в слабых ферромагнетиках было обнаружено новое явление, в котором, по-видимому, также может проявляться эффект туннелирования доменных границ. При исследовании высокоскоростной динамики границ в слабых ферромагнетиках $YFeO_3$ и $TmFeO_3$, был обнаружен эффект потери границей устойчивости, проявлявшийся в существенно нестационарном движении [38]. Такое состояние характеризовалось формированием двойных динамических доменных структур, и одновременным существованием границ, движущихся с различными скоростями. Автор [38] предположил, что возникновение двойников связано с туннелированием границ через периодически расположенные в кристалле слабого ферромагнетика магнитные неоднородности. Однако точность выполненных к настоящему времени измерений недостаточна для оценки высоты потенциальных барьеров таких неоднородностей. Поэтому сложно вы-полнить соответствующие оценки для характеристик туннельных процессов и, следовательно, нельзя с уверенностью связывать обнаруженные динамические аномалии с туннелированием.

---

[14]Заметим, что наличие электронов проводимости, вообще говоря, может вызвать интерференционные эффекты, создающие каналы для диссипации. Снизить влияние этого эффекта в эксперименте можно либо увеличивая размеры доменных границ, либо подбирая материалы с низкой электропроводностью. Туннелирование в присутствии диссипации мы обсуждаем в главе 4.

## 2.3 Туннелирование. Точные решения и ВКБ — приближение

В теории туннелирования границ в ферромагнетиках можно использовать для вычислений четыре основных метода:
— 1) прямое решение уравнения Шредингера для заданного потенциала;
— 2) классичесий ВКБ-метод;
— 3) инстантонную технику;
— 4) численные методы. Каждый из названных методов имеет как свои преимущества, так и недостатки.

### 2.3.1 Точные решения уравнения Шредингера

Преимущества метода очевидны. К недостаткам можно отнести ограниченность класса потенциалов, для которых может быть получено уравнение Шредингера в явном виде. Тем не менее, поскольку существует некоторая свобода выбора в моделировании формы потенциальных барьеров, возникающих для доменных границ в кристаллах, можно уже на этапе моделирования постараться свести задачу к известному типу потенциала. Выше (п. 1.1.4, выражение (38)) мы получили выражение для потенциала взаимодействия блоховской границы с точечным дефектом. Для этого потенциала задача может быть решена точно (см. [41], задача 4 к пар. 25, [43]). В случае отталкивающей примеси $U_0 > 0$ стационарное уравнение Шредингера имеет вид

$$\frac{d^2\psi}{dx^2} + \frac{2m}{\hbar^2}\left(E + \frac{U_0}{\cosh^2(x/D)}\right)\psi = 0. \quad (48)$$

Решение (48) имеет вид

$$\psi = (1-\xi^2)^{-ikD/2} \times$$
$$\times F\left(-ikD-s, -ikD+s+1, -ikD+1, \frac{1-\xi}{2}\right), \quad (49)$$

где

$$\xi = \tanh(x/D),$$
$$k = \frac{1}{\hbar}\sqrt{2mE},$$
$$s = \frac{1}{2}\left(-1 + \sqrt{1 - \frac{8mU_0D^2}{\hbar^2}}\right).$$

Соответствующий же коэффициент прохождения будет

$$\mathcal{D} = \frac{\sinh^2 \pi kD}{\sinh^2 \pi kD + \cos^2\left(\frac{\pi}{2}\sqrt{1 - \frac{8mU_0D^2}{\hbar^2}}\right)}$$
$$\text{для } \frac{8mU_0D^2}{\hbar^2} < 1, \quad (50)$$

$$\mathcal{D} = \frac{\sinh^2 \pi kD}{\sinh^2 \pi kD + \cosh^2\left(\frac{\pi}{2}\sqrt{\frac{8mU_0D^2}{\hbar^2} - 1}\right)}$$
$$\text{для } \frac{8mU_0D^2}{\hbar^2} > 1. \quad (51)$$

В разделе 1.1.1 мы уже замечали, что в большинстве известных экспериментов параметры границы имеют такие величины, которые делают вероятность туннелирования, рассчитанную в традиционном подходе исчезающе малой. С помощью выражений (50) и (51) легко оценить абсолютную величину коэффициента прозрачности потенциального барьера. На рис. 1.9 и 1.10 мы приводим результаты вычисления $\ln \mathcal{D}$ как функции энергии $E$ границы для различных значений толщины $D$ и эффективной массы $m$ туннелирующего участка границы. Из рисунков видно, что для реальной регистрации туннелирования необходимо, чтобы либо границы были по крайней мере на порядок уже, либо, чтобы их эффективная масса была меньше на один - два порядка. Последнего можно добиться опять же либо уменьшением дерринговской массы, либо уменьшением площади туннелирующего участка границы. Однако ни тот, ни другой вариант, по-видимому, нельзя использовать при анализе уже известных экспериментов. Поэтому вновь подчеркнем, что очень актуально рассмотрение альтернативных физических механизмов, приводящих к более реалистичным оценкам.

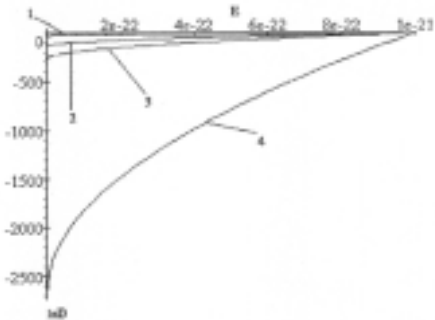

Рис. 9. Коэффициент прозрачности $\mathcal{D}$ в логарифмической шкале, как функция энергии границы $E$ для различных значений толщины границы $D$: 1 - $D = 10^{-7} cm$, 2 - $D = 5 \cdot 10^{-6} cm$, 3 - $D = 10^{-6} cm$, 4 - $D = 10^{-5} cm$. Расчеты выполнены для значения массы $m = 10^{-25} g$.

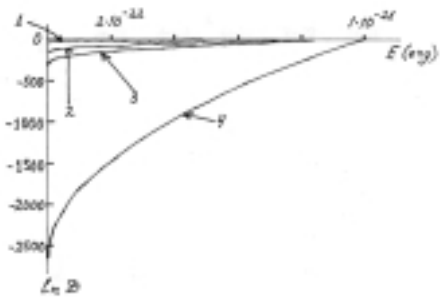

Рис. 10. Коэффициент прозрачности $\mathcal{D}$ в логарифмической шкале, как функция энергии границы для различных значений эффективной массы границы $m$: 1 - $m = 10^{-26} g$, 2 - $m = 2 \cdot 10^{-26} g$, 3 - $m = 6 \cdot 10^{-26} g$, 4 - $m = 10^{-25} g$.

### 2.3.2 Вычисления с использованием ВКБ-метода

Существо метода хорошо известно и мы не будем здесь на нем подробно останавливаться. К недостаткам метода, очевидно, нужно отнести то обстоятельство, что он не позволяет вычислять предэкспоненциальный множитель в туннельной экспоненте. Однако, во-первых, в большинстве случаев точность получаемых из экспериментальных данных параметров не очень высока, и в задаче требуется лишь определить качественное согласие эксперимента и теории, что делает использование ВКБ-приближения вполне уместным. Во-вторых же, известны задачи, для которых ВКБ-приближение дает точное решение. Ниже мы будем неоднократно обращаться к рассмотрению частицы в поле перевернутого параболического потенциала $U(x) = -\frac{1}{2}\kappa x^2$. Задача о туннелировании в этой ситуации была решена Кемблем в 1935 г. [42], причем подчеркнем, что полученное решение для вероятности туннелирования является точным. Коэффициент прохождения дается выражением

$$\mathcal{D} = \frac{1}{1 + e^{-2\pi\epsilon}}, \qquad (52)$$

где

$$\epsilon = \frac{E}{\hbar}\sqrt{\frac{m}{\kappa}}.$$

Если частица имеет отрицательную энергию (максимум потенциала здесь соответствует нулевой энергии), и велика по абсолютной величине, то выражение (52) переходит в $\mathcal{D} \cong e^{-2\pi\epsilon}$. Для положительных $E$,

$$\mathcal{R} = 1 - \mathcal{D} = \frac{1}{1 + e^{2\pi\epsilon}}$$

есть вероятность надбарьерного отражения.

## 2.4 Туннелирование в эвклидовом времени. Инстантоны

Достаточно популярным в последние годы стал метод решения квантово-механических задач с использованием интегралов по траекториям. В главе

3 мы продемонстрируем как в этой технике может быть решена задача о термостимулированном туннелировании. Здесь же остановимся на обсуждении простых примеров, которые демонстрируют основные особенности метода. Кроме того, мы будем возвращаться к этим примерам в ходе дальнейшего изложения, поскольку в некоторых случаях они будут использоваться для построения конкретных вычислительных моделей.

Рассмотрим частицу массы $m$, движущуюся в независящем от времени потенциале $U(x)$. Амплитуда перехода такой частицы из точки $x$, где она находилась в момент времени $t = 0$ в точку $y$ к моменту времени $t$ дается функцией Грина

$$G(x,y,t) = \langle y \left| \exp\left(-iHt\right) \right| x \rangle, \quad (53)$$

где $H$ — гамильтониан системы. Фейнман и Гиббс [45] предложили для вычисления амплитуды перехода использовать так называемые интегралы по траекториям

$$G(x,y,t) = \int Dx(t) \exp(iS[x(t)]). \quad (54)$$

В таком формализме функция Грина дается суммой по всем возможным путям $x(t)$, ведущим из $x$ при $t = 0$ в $y$ к моменту врмени $t$, при этом вес траекторий дается действием

$$G(x,y,t) = \int Dx(t) \exp(iS[x(t)]). \quad (55)$$

Более аккуратно можно дать определение интеграла по траекториям, вводя дискретизацию траекторий. Разбивая временную ось на $N$ интервалов шириной $a = t/N$, можно представить интеграл по траекториям как $N$-мерный интеграл по $x_n = x(t_n = an)$, $n = 1, \ldots, N$. Дискретизированное действие тогда дается как

$$S = \sum_n \left[ \frac{m}{2a}(x_n - x_{n-1})^2 - aU(x_n) \right]. \quad (56)$$

Интеграл по траекториям сводится, таким образом, к кратному интегралу по $x_n$, где $n \to \infty$. Вообще говоря, в явном виде могут быть вычислены лишь интегралы гауссова типа. Например, для гармонического осциллятора $U(x) = \frac{m\omega^2 x^2}{2}$ [45]

$$G_{osc} = \left( \frac{m\omega}{2\pi i \sin \omega t} \right)^{1/2} \times$$
$$\times \exp\left[ \left( \frac{im\omega}{2\sin \omega t} \right) \left((x^2 + y^2)\cos \omega t - 2xy\right) \right]. \quad (57)$$

В то же время, действие (56) может быть найдено и численными методами. В этом случае, однако, возникает проблема с вычислением сильно флуктуирующих фаз. Обойти эту сложность можно следующим путем. Нужно выполнить аналитическое продолжение функции Грина на мнимую (эвклидову) временную ось: $\tau = it$. Тогда весовая функция принимает форму $\exp(-S_E[x(\tau)])$, где $S(E)$ - эвклидово действие

$$S = \int d\tau \left[ m\frac{\dot{x}^2}{2} + U(x) \right]. \quad (58)$$

В этом случае вес оказывается положительно определенным и численное моделирование вполне оправдано.

Подобным приемом перехода во мнимое (эвклидово) время можно воспользоваться и при анализе туннелирования. В самом деле, энергия частицы, движущейся в потенциале $U(x)$, есть

$$E = \frac{p^2}{2m} + U(x).$$

С точки зрения классической механики доступной является только та область фазового пространства, где кинетическая энергия положительна. В классической запрещенной области кинетическая энергия отрицательна, что соответствует мнимому импульсу $p$. Однако, в квантовом случае рассмотрение волновой функции в классически недоступной области является вполне осмысленным. Например, волновая функция частицы в ВКБ приближении есть

$$\psi(x) \sim \exp\left[i\Phi(x)\right],$$

где

$$\Phi(x) = \pm \int_0^x dx' p(x') + O(\hbar)$$

и
$$p(x) = \sqrt{2m\left[E - U(x)\right]}.$$

В классически доступной области волновая функция — гармоническая, тогда как в классически запрещенной области (соответствующей мнимому импульсу) она экспоненциально подавлена. Выполняя аналитическое продолжение $\tau = it$, классическое уравнение движения можно записать в виде

$$m\frac{d^2 x}{d\tau^2} = +\frac{dV}{dx}. \qquad (59)$$

В (59) знак потенциальной энергии изменился, что делает недоступную область во мнимом времени доступной в классическом смысле. Выдающаяся роль классического пути туннелирования становится понятна при рассмотрении фейнмановских интегралов по траекториям. А именно:

*Несмотря на то, что в квантовой механике допустимы любые траектории, интеграл по траекториям доминирует на тех путях, которые максимизируют весовой фактор $\exp(-S[x_{cl}(\tau)])$, или минимизируют эвклидово действие.*

Классическая траектория — это траектория с минимально возможным действием.

К понятию инстантона проще всего прийти, рассматривая задачу о туннелировании в двухуровневой системе между метастабильными вакуумными состояниями. Такая задача моделирует ситуацию, например, в молекулярных магнитных кластерах, или в однодоменных ферромагнитных частицах [46], [47]. Форма потенциала представляет собой двойную симметричную яму с минимумами, расположенными, скажем, в точках $-\eta$ и $\eta$. Пусть потенциал определяется функцией

$$U(x) = \lambda(x^2 - \eta^2)^2. \qquad (60)$$

Из закона сохранения энергии имеем

$$\frac{m\dot{x}^2}{2} - U(x) = 0. \qquad (61)$$

Решением (61) является функция

$$x_{cl}(\tau) = \eta \tanh\left[\frac{\omega}{2}(\tau - \tau_0)\right], \qquad (62)$$

причем $\omega^2 = \frac{8\lambda\eta^2}{m}$. Классическое действие, соответствующее (62) будет

$$\int_{-\infty}^{\infty} d\tau \left[\frac{m\dot{x}^2}{2} + U(x)\right] = \frac{\omega^3\lambda}{12m}. \qquad (63)$$

*Траектория, определяемая (62), идущая из $x(-\infty) = -\eta$ в $x(\infty) = \eta$ называется инстантонной траекторией или инстантоном.*[15]

Амплитуда туннелирования в нулевом приближении здесь будет

$$P \sim e^{-S_0}.$$

Высшие порядки квазиклассического приближения для интеграла по траекториям могут быть получены разложением вокруг классического решения

$$\langle -\eta | e^{-H\tau} | \eta \rangle =$$
$$= e^{-S_0} \int Dx(\tau) exp\left(-\frac{1}{2}\delta x \frac{\delta^2 S}{\delta x^2}_{xcl} \delta x + \ldots\right), \quad (64)$$

т.е. вычисление предэкспоненциального множителя требует расчета флуктуаций вокруг классического инстантонного решения.

Напомним, что до сих пор мы рассматриваем туннелирование между метастабильными вакуумами системы. Однако при переходе к изучению кинетики процессов срыва необходимо обобщить задачу применительно к туннелированию из возбужденных состояний. В этом случае, переход из начальной точки в конечную происходит, вообще говоря, за конечное время. Это обстоятельство можно учесть вводя в рассмотрение периодические инстантоны по схеме, описанной ниже.

Рассмотрим туннелирование сквозь перевернутый параболический потенциал $U(x) = -\kappa x^2/2$ (обсуждавшаяся выше задача Кембля [42]). Для частицы с энергией $E < 0$ область между точками $x_{1,2} = \pm\sqrt{\frac{2|E|}{\kappa}}$ классически недоступна, однако после перехода во мнимое время получается вещественное действие

$$S_0 = \int_0^T E dt = E \cdot 2\pi\sqrt{m/\kappa}, \qquad (65)$$

---
[15]По аналогии с солитоном, локализованным в пространстве, тогда как инстантон оказывается локализованным во времени.

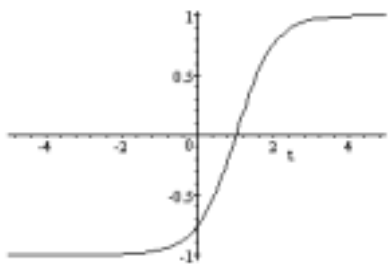

Рис. 11. $\tau_0$ определяет положение центра инстантона. На рисунке $\tau_0 = 1$, $\eta, m, \omega = 1$.

а амплитуда туннелирования в нулевом порядке будет
$$P(E) \sim e^{-\frac{S_0}{\hbar}},$$
что полностью соответствует решению Кембля для больших по модулю значений $E$.

Так же, как и для классических инстантонов, высшие порядки в разложении решения получаются расчетом флуктуаций вокруг основного решения [16].

## 2.5 Итоги

В обзоре рассмотрены наиболее часто используемые модели, описывающие магнитодинамику в магнитных материалах разных типов — ферро, антиферро-, слабых ферромагнетиках и др., и сформулированы математические задачи, соответствующие этим моделям.

Подробно рассмотрено взаимодействие мезоскопических объектов — доменных границ — с магнитными неоднородностями среды. Показано, как это взаимодействие сказывается на макроскопических характеристиках процессов перемагничивания. Установлено, что учет квантовых и термических процессов срыва доменных границ требует пересмотра определения макроскопических магнитных характеристик, таких как коэрцитивная сила.

---

[16] Более подробно периодические инстантоны мы рассматриваем в главах 3 и 4.

Установлено, что в большинстве случаев, проявление квантовых эффектов, например, туннелирования, или надбарьерного отражения, доменных границ, само по себе имеет крайне низкую вероятность. В то же время, на основе сопоставления с известными экспериментальными результатами других авторов, делается вывод о возможности возникновения подобных эффектов в комплексе с другими процессами. Например, возможно проявление срыва границ, как результат термостимулированного туннелирования, или резонансного туннелирования при воздействии внешнего периодического возмущения. Детальные результаты исследования таких процессов мы приведем в следующих публикациях.

## Список литературы